\documentclass[
    ,final            
  ]
  {aipproc}

\layoutstyle{6x9}

\begin{document}

\title{Progress in measuring neutrino quasielastic interactions}


\author{Richard Gran}{
  address={University of Minnesota Duluth}
}



\begin{abstract}
This is an exciting time for folks who are looking at neutrino
cross sections, and the especially important quasielastic interaction.  
We are able to inspect several recent results
from K2K and MiniBooNE and are looking forward to a couple more
high statistics measurements of neutrino and anti-neutrino 
interactions.  There is additional interest because of the need
for this cross section information for current and upcoming
neutrino oscillation experiments.  This paper is a brief review
of our current understanding and some puzzles when we compare 
the recent results with past measurements.  I articulate
some of the short term challenges facing experimentalists, 
neutrino event generators, and theoretical work on the 
quasielastic interaction.
\end{abstract}

\maketitle


\section{Introduction}
As this century's neutrino oscillation experiments have produced their first results,
so have the associated neutrino interaction measurement efforts.
The K2K\cite{K2K-SciFi,K2K-SciBar} 
and MiniBooNE\cite{MiniBooNE} 
experiments have presented their investigations of the charged-current quasielastic 
(CCQE)
neutrino cross section $\nu_\mu + n \rightarrow \mu^- + p$ in 
carbon and oxygen.  MiniBooNE has additionally shown its first distributions
for the anti-neutrino cross section.  Their initial results seem to be at odds with 
the previous measurements from deuterium bubble chamber experiments.

Upcoming neutrino oscillation experiments depend heavily on the 
quasielastic interaction.  It is a very large portion of the 
event rate around 1 GeV where oscillations are expected and where
the current neutrino beams are tuned.  Also, 
the two-body kinematics of this interaction are vital for obtaining
an estimate of the incident neutrino energy spectrum for those experiments
that depend on Cerenkov light, such as MiniBooNE and SuperK.
An apparent puzzle with this cross section is a concern,
whether it is due to cross section, nuclear, or experimental systematic
effects.

\section{Status of experimental results}

\begin{figure}
  \includegraphics[height=.5\textheight]{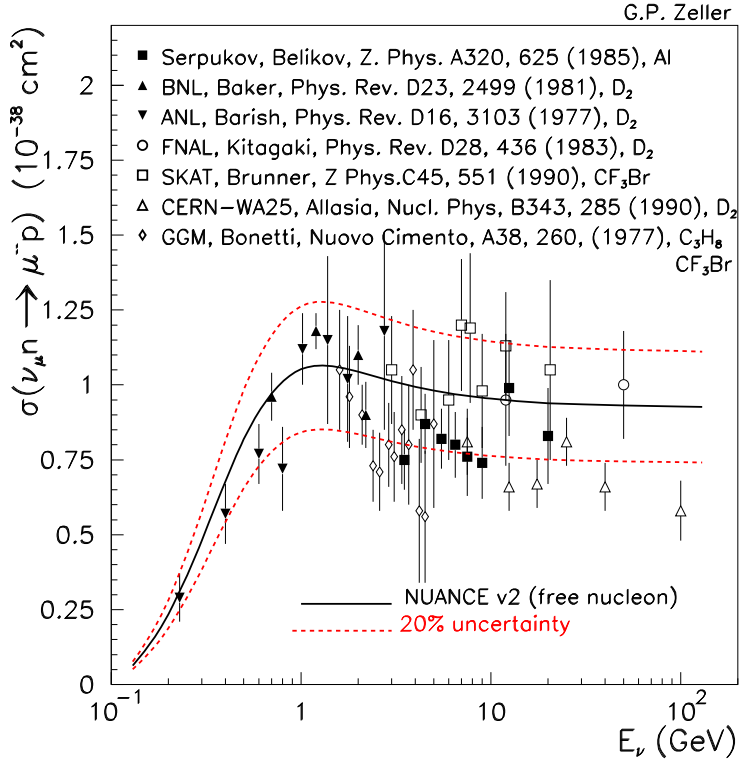}
  \caption{Summary of most of the pre-2000 data on neutrino CCQE interactions,
plot courtesy of G. Zeller\cite{zeller}.
The data from BNL, ANL, and FNAL, are based 
significantly on a shape
fit to the $Q^2$ distribution, the others use a determination based
on their rate and flux, or a mix of rate and shape.}
  \label{fig.qezeller}
\end{figure}

There are two kinds of CCQE measurements.  The direct measurement of a cross section is
based on the measured rate divided by the flux and the number of targets,
usually expressed as $\sigma_{QE}(E)$.  
It is a difficult measurement 
to make because we are challenged to improve our knowledge of the neutrino flux in our experiments.
The alternative is an analysis of the shape of the $Q^2$ distribution which 
can be less dependent on the flux uncertainties.  Such results assume the
vector form factors are given by electron scattering experiments and extract information
about the axial vector form factor, most often assuming a dipole shape with one 
free parameter, the axial vector mass $M_A$. 
Using the formalism summarized by Llewellyn-Smith\cite{LlewellynSmith}
and the extracted form factors, we obtain information about the total 
cross section, though this is a somewhat more model dependent statement.
The previous world measurements of the quasielastic cross section have
been summarized in plots such as Fig.~\ref{fig.qezeller}.  

An important consideration in this plot is that it contains a mix of 
the two different techniques, and the bubble chamber measurements at the 
lowest energies (the BNL and ANL data) 
are from fits to the shape of the $Q^2$ distribution.
These lowest energy measurements are also the measurements that had 
the smallest errors on the extracted value of $M_A$ and dominated the
global fits for $M_A$ that were used in neutrino event
generators until recently.  

A sequence of papers that provide excellent lesson on this point are
from the Argonne National Lab (ANL) experiment\cite{ANL}; their publications contain
results using four methods with different dependencies 
on the uncertainty in their neutrino flux.  In that case, the estimated cross section 
comes out 20\% lower (approximately one sigma); the errors are dominated
by the uncertainty in the flux.  The $\sigma(E)$ points used in the plot above
are based on an analysis that uses their flux-independent shape of the $Q^2$
to get an $M_A$, then re-extract a flux.  This new flux is averaged with their
default flux to produce those data points.
Ideally, both techniques would produce compatible results.
We still understand the ANL authors' conclusions: the  tension between these two results 
can be understood to mean that the uncorrected flux estimate was too high or the 
shape of the $Q^2$ distribution and the axial form factors does not map
so simply to the total $\sigma(E)$.  One sigma is not a large discrepancy,
but an oscillation experiment must assign an error and mitigate the
effects of a large uncertainty on $\sigma(E)$, rather than the smaller errors 
on the shape of $d\sigma/dQ^2$.

\subsection{The new results}

I summarize the newest results very quickly, more detailed information
is available elsewhere in these proceedings or in their publications.  
There are three new determinations of $M_A$ from the shape of the $Q^2$
distribution, all of which are higher than the bubble chamber results.
K2K reports a value of 1.20 $\pm$ 0.12 from the SciFi detector\cite{K2K-SciFi} using
a mixture of water and aluminum as a target, and also a preliminary result
of 1.14 $\pm$ 0.11 from the SciBar detector\cite{K2K-SciBar} using a scintillator (CH) target.
These two experiments share some systematic errors in the beam flux and 
muon momentum determination, and have not yet given guidance about
whether the results should be combined.  

The MiniBooNE experiment has 
added a very high statistics measurement and also investigated a way
of parameterizing the very low $Q^2$ region\cite{MiniBooNE}.  They obtain a value for 
$M_A$ = 1.23 $\pm$ 0.20 using this parameterization and 1.25 $\pm$ 0.12
when repeating their analysis with only $M_A$ as a parameter and not
including the very low $Q^2$ region in their fit, similar to the technique
used by K2K and the bubble chamber experiments.  That these results are 
higher is saying that there are relatively more events at high $Q^2$
compared to the default Monte Carlo simulation predictions.

The MiniBooNE result is
noteworthy for several reasons.  It is the first of what I will call an ``enormous statistics'' 
measurement, which will give them other avenues to understand systematic
effects. For example they have shown their raw $p_\mu$ vs. $\theta_\mu$ plot,
in addition to the reconstructed $Q^2$ distribution.  They also emphasize 
the essential need to understand the very low $Q^2$ region for their 
oscillation analysis; a description of this region had been neglected 
up to this point.

In the very near future, we expect to hear more information on quasielastic reactions
from several sources.  The NOMAD experiment expects to have a final QE result
on a modestly short timescale.  There is a little new information 
on the QE-like event rate in the emulsion
in the CHORUS experiment\cite{chorus}.  The analysis of the MINOS data is currently
underway; they are sitting on the next enormous statistics dataset.
MiniBooNE has started showing distributions from anti-neutrino
running, and is also expected to provide information using the rate/flux 
method.  They state that their flux is constrained to 15\% via the HARP
hadron production measurements and other analyses.

On a somewhat longer timescale we expect results from SciBooNE, the SciBar detector
operating in the MiniBooNE beam, which has already taken a significant amount
of anti-neutrino data.  After that will come yet another enormous statistics 
analysis from the MINERvA experiment, including a systematic investigation
of the QE reaction on different nuclear targets CH, He, C, Fe, and Pb.
The MINERvA and MINOS measurements are on different
nuclei and a different energy range than the K2K and MiniBooNE results.\
The MINOS experiment now has and the MINERvA experiment expects to have 800,000
quasielastic interactions with energies between 1 and 20 GeV, 
as predicted by their Monte Carlo and estimates of their run plan.

\section{Upcoming experimental challenges}

\subsection{Systematic Errors}
The reconstructed $Q^2$ distribution is affected by several systematic
errors that are a challenge to control and which have overlapping effects
on the analysis and final errors.  If the primary focus of a shape fit is
the $Q^2$ distribution, then the shape of the beam flux, the size of the
bias in the muon momentum reconstruction, and the $Q^2$ shape as parameterized
with $M_A$ have interlocking correlations.
I will ignore systematics related to event
selection and resolution, which are too experiment specific to discuss here.  

\begin{figure}
  \includegraphics[width=.53\textwidth]{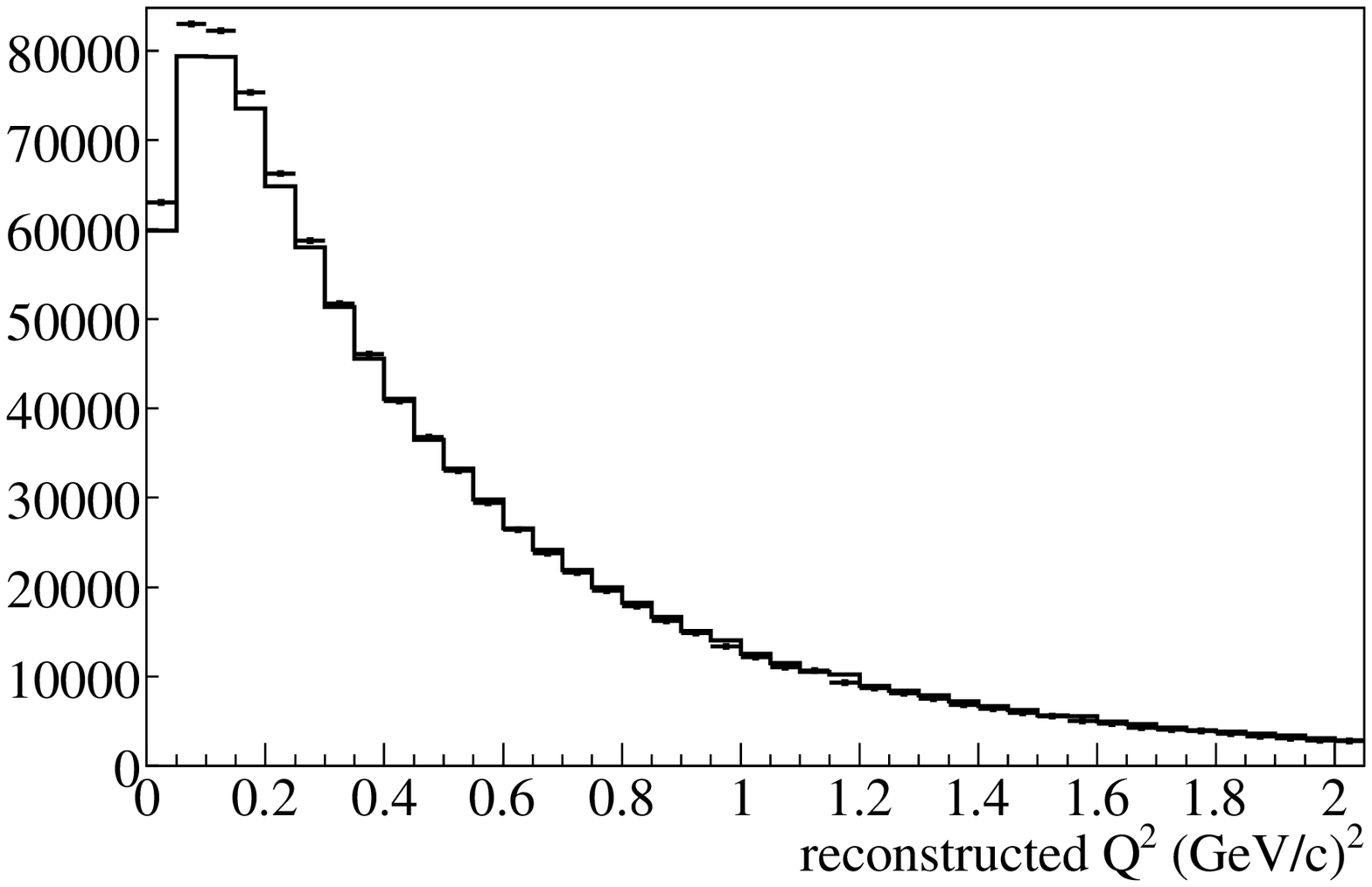}
  \includegraphics[width=.53\textwidth]{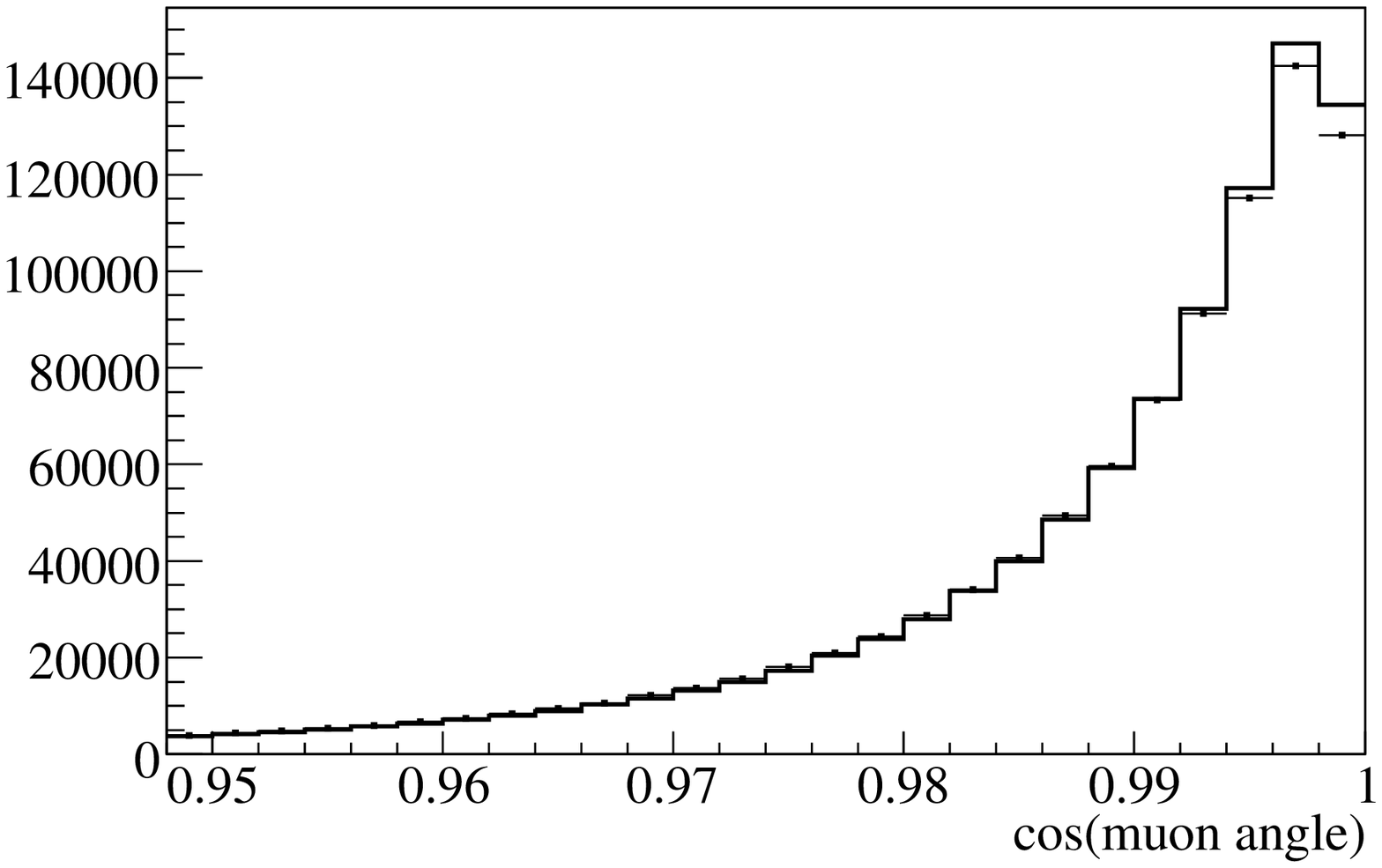}
  \caption{Two plots illustrating the effect of two systematic biases.
Both are from a {\sc neugen}\cite{neugen} calculation of $\nu_\mu$ interactions on 
carbon.  The nominal distribution is shown with the line.  The points
in the left plot of $Q^2$ have a -2\% bias in $p_\mu$, while the 
right plot of $\cos\theta_\mu$ has a -2\% bias in $E_\nu$.  The statistical
errors are negligible in both cases.
}
  \label{fig.systematics}
\end{figure}

The left plot of Fig.~\ref{fig.systematics} shows the distortion of the
$Q^2$ spectrum produced by a muon momentum bias.  For this mono-energetic
5.0 GeV sample of neutrino-carbon interactions, the points 
are reconstructed with a -2.0\% $p_\mu$ bias,
relative to the histogram.  If this was the only significant error, 
the extracted $M_A$ would be biased between 5\% and 10\%.
The raw $p_\mu$ and $Q^2$ would show this discrepancy while
the $\theta_\mu$ would be well reproduced, because this bias shouldn't 
have any effect on the angle reconstruction.  

This bias can be caused by
the detector material assay, magnetic field errors, optical model, track vertex or end 
biases, which are specific to each experiment.
There may also be some latent bias in the {\sc geant3} or {\sc geant4} muon dE/dx
simulation, depending how the experiment calibrates their momentum reconstruction.
Especially for the lower energy beams such as used by MiniBooNE, SciBooNE, and the K2K
near detectors, errors in modeling the (beyond the) Fermi gas removal energy 
parameters or the Coulomb interaction of the outgoing charged lepton could 
play a small role equivalent to an apparent $p_\mu$ bias.

The situation would be simple, except that another source of error has similar 
effects.  The beams we use are usually peaked at some value, and the location
of that peak depends significantly on the primary beam targeting and the 
hadron focusing optics; an illustration of some uncertainties involved can be
found in \cite{NuMI:2006}.  The right-side plot 
in Fig.~\ref{fig.systematics} 
is a simplified demonstration of this; the mono-energetic beam is shifted
low by 2\% for the points.  This plot shows the distortion of the
$\cos\theta_\mu$ distribution.  There is a trivial 2\% distortion of 
the $p_\mu$ distribution as well, but the $Q^2$ distribution comes out 
undistorted.  Note, this is actually a small bias in the shape of 
the underlying flux, relative to the width of the peak, 
typical $E_\nu$ binning, and resolution of these experiments.

Extracting $M_A$ is now a challenge if the external constraints on the
pair of systematic parameters are not very strong.  If one or the other 
is not negligible, the analysis
technique should attempt to incorporate both kinds of errors into the $M_A$
fitting.  The two published results \cite{K2K-SciFi,MiniBooNE} have expressed
different attempts to deal with this.  The new and upcoming very large datasets
may allow for more sophisticated ways of dividing up $p_\mu$ and $\theta_\mu$
in such a way that these systematics can be convincingly isolated with the
neutrino data themselves.

\subsection{Rate and Shape measurements}

In my opinion, the other major task facing the experimentalists is to 
provide both rate and shape measurements of the quasielastic cross section.
As discussed earlier in this paper, some experiments have done this 
in the past and reported slightly more than one-sigma discrepancies,
when the large errors on the flux are taken into account.  The BooNE and
NuMI beams at Fermilab appear to offer the possibility of a flux constraint
at least as good as was achieved in the past using the very large neutrino
datasets and the existing beam instrumentation.
Initiatives and upgrades to the NuMI instrumentation, 
driven by the MINOS and MINERvA experiments,
may yield substantial improvement over this, and give better confirmation
that the QE cross section model and $M_A$ can adequately describe 
the $Q^2$ shape and the interaction rate at the same time.

\section{Upcoming theory challenges}

\subsection{Beyond Fermi gas and attention to reconstructed quantities}

There is general concern that the implementation of nuclear effects in 
the current neutrino event generators is inadequate.  These enormous
statistics data sets are likely to reveal more deficiencies than in the past,
and there are several
efforts underway to incorporate better models.  
This includes spectral function and related
models which take into account correlations between nuclei in the nuclear
target, and also improved final state hadron rescattering models.

\begin{figure}
  \includegraphics[width=.50\textwidth]{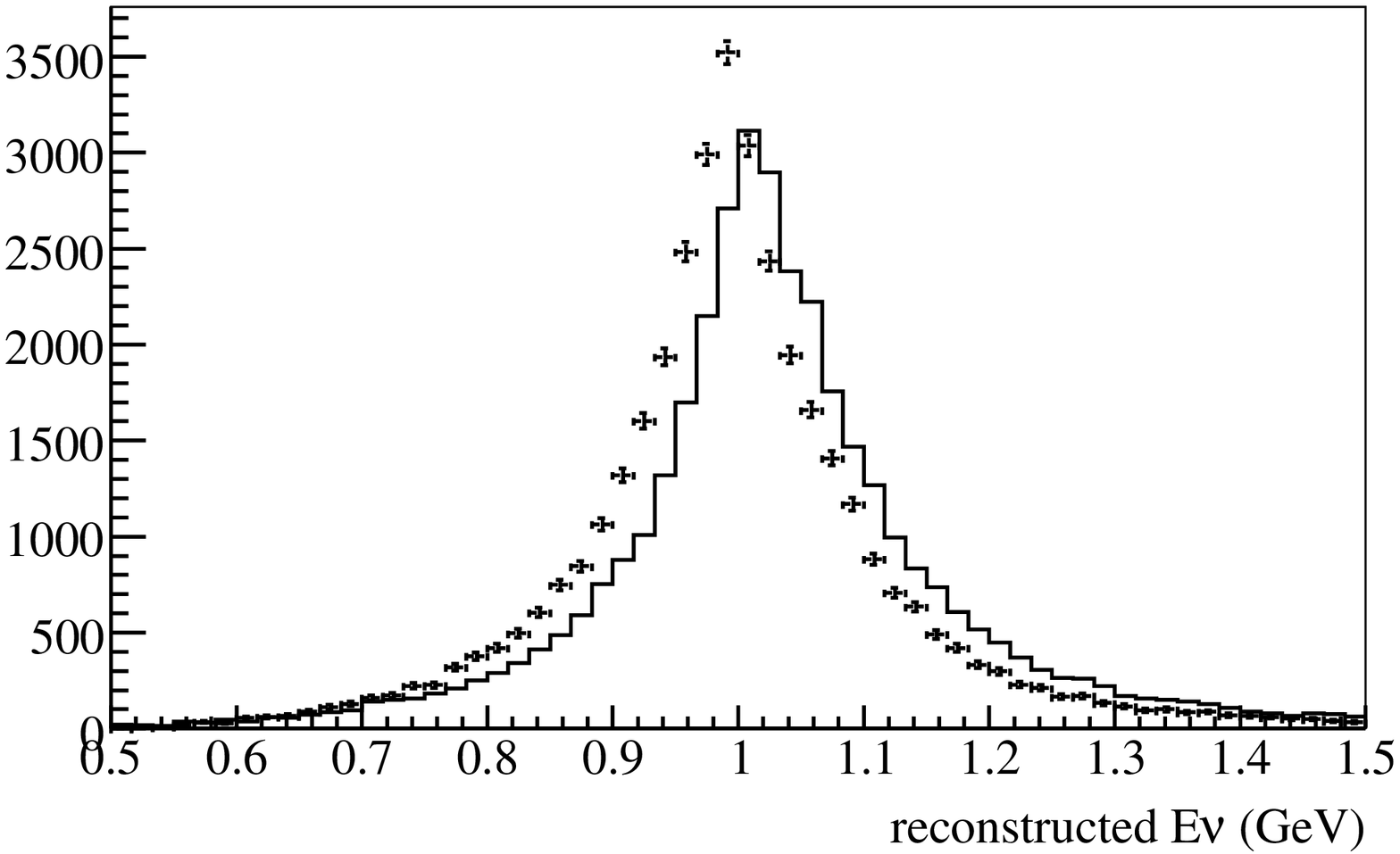}
  \includegraphics[width=.50\textwidth]{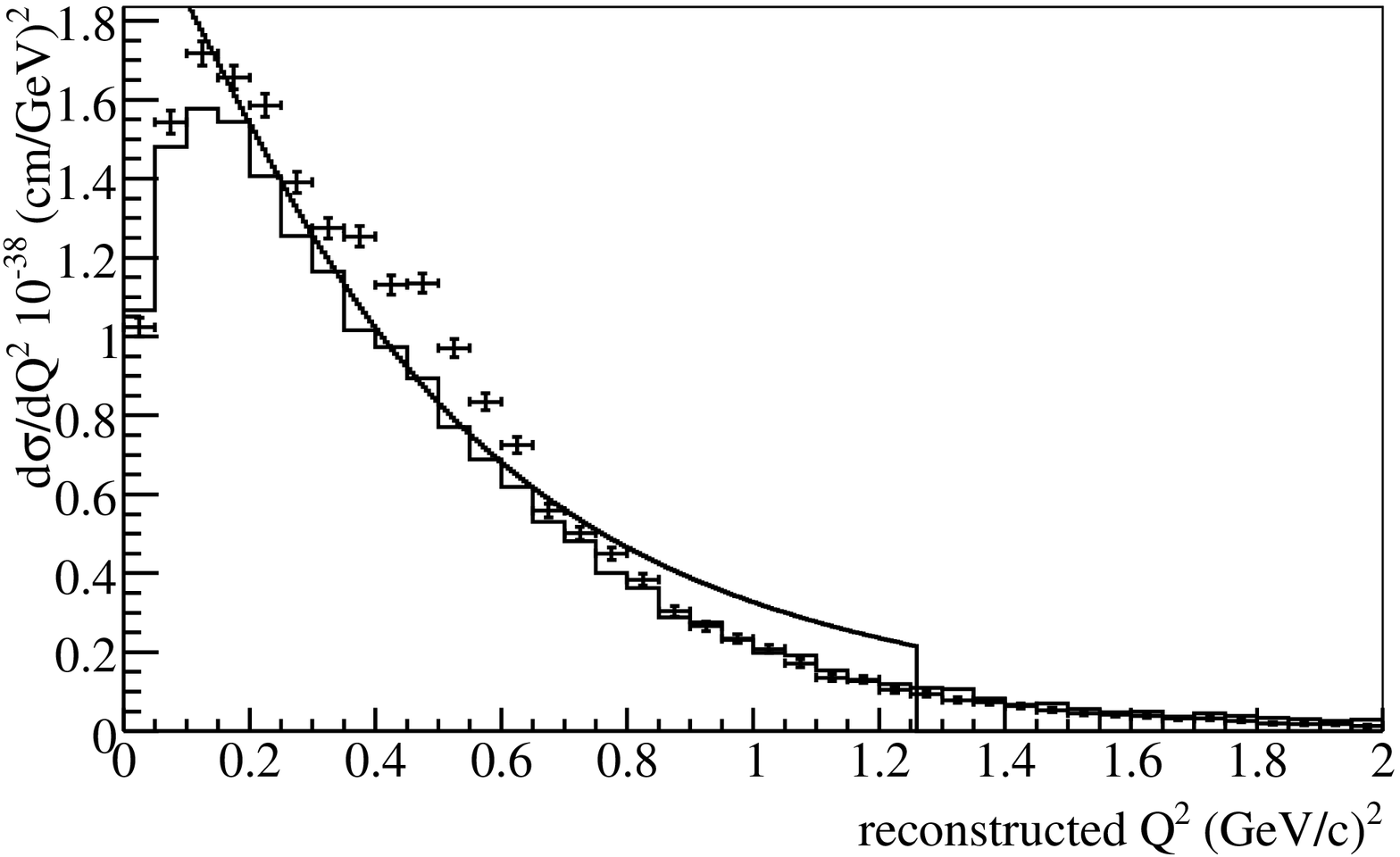}
  \caption{A comparison of the reconstructed $Q^2$ and $E_\nu$ quantities
for the {\sc neugen} (solid histogram) and {\sc neut} (points with statistical errors)
event generators.  The $Q^2$ plot has the distribution scaled to represent
the effective cross section for a single one of the neutrons in oxygen, and 
the additional solid line shows the corresponding cross section for 
interactions with a single free neutron.  The reasons for the 
visible distortions in these spectra are described in the text.
The statistical errors for {\sc neugen} are similar to the ones shown for {\sc neut}.
}
  \label{fig.recoquantities}
\end{figure}

As these new calculations become available, it is important to take
note of the definition of reconstructed $Q^2$ and $E_\nu$ being used
to report the experimental results.  There are additional distortions
to these reconstructed spectra, above what would be expected from
the techniques familiar from electron scattering.
This definition starts with the usual expression
$${\mathrm{reconstructed}}\;Q^2 = 2 E_\nu (E_\mu - p_\mu cos\theta_\mu) - m_\mu^2.$$
A major difference from current theory calculations and experience
from electron scattering is that the value of the neutrino 
energy $E_\nu$ is not known on an event by event basis.
In this case, the neutrino energy can be estimated with the following
expression
$${\mathrm{reconstructed}}\;E_{\nu} = \frac{(m_N + \epsilon_B)E_\mu -
                  (2m_N \epsilon_B + \epsilon_B^2 + m^2_\mu)/2}
                  {m_N + \epsilon_B - E_\mu + p_\mu \cos \theta_\mu}$$
where $\epsilon_B \approx -27$ MeV for Oxygen 
is the removal energy and here I take it to be intrinsically
negative.  This expression is then used in the $Q^2$ calculation.  
This reconstructed $E_\nu$ 
is smeared and possibly biased because of the the nucleon momentum distribution, even 
before considering detector resolution effects; a comparison
of the smearing from {\sc neugen}\cite{neugen} and {\sc neut}\cite{neut}
models is in the left hand plot of Fig.~\ref{fig.recoquantities}.
This use of a reconstructed $E_\nu$ introduces one of several distortions
into the reconstructed $Q^2$ spectrum, shown in the right hand plot of 
Fig.~\ref{fig.recoquantities}.  A related expression can be used if
the event likely produced a 1232 MeV resonance.

These plots show the reconstructed $E_\nu$
and $Q^2$ distributions for 1.0 GeV neutrinos incident on oxygen from 
the {\sc neugen} (solid histograms) and {\sc neut} (points with statistical errors)
neutrino event generators, with no detector effects or backgrounds.  
The right hand plot has been scaled to represent the cross section 
for a neutrino incident on just one of the eight neutrons in oxygen,
though with no unfolding of the distortions.
The solid line in the $Q^2$ plot is the free neutron cross section for
a very similar set of model parameters as {\sc neugen}.

There are several causes for the visible distortions in that $Q^2$ spectrum.
The choice of $M_A$ = 1.1 GeV for {\sc neut} increases the cross section
by about 10\%, compared to {\sc neugen} and the given free neutron line.
Pauli blocking is implemented for both event generators, suppressing the
very low $Q^2 < 0.2$ (GeV/c)$^2$ region.  The kinematic cutoff just 
above $Q^2 = 1.25$ (GeV/c)$^2$ is smeared out by the neutron Fermi motion.
These are relatively large effects.

But there are more interesting things in this comparison, which need
to be addressed with more careful calculations of the nuclear environment.
The {\sc neugen} code implements a tail to the Fermi motion distribution
that goes up to 500 MeV/c based on the paper by Bodek and Ritchie\cite{bodek}.
Both event generators have a peak structure near $Q^2 = 0.5$ (GeV/c)$^2$ 
when using these reconstructed quantities,
but this structure is smeared out for {\sc neugen} because of this tail
and is not as prominent.
Other aspects
of a beyond-the-Fermi gas distribution would have different, potentially
observable distortions of this reconstructed $Q^2$ spectrum that would not
be apparent in a simpler $Q^2$ spectrum.  A related question is what 
removal energy is appropriate for this reconstruction, when an experiment
applies it to both data and Monte Carlo?  The removal energy, off mass shell 
nature of the nucleon, and their effects on the resulting lepton kinematics
should be inspected, and may contribute to shifts in the reconstructed $E_\nu$
spectrum and distortions of the reconstructed $Q^2$ spectrum.

In addition to this reconstruction $Q^2$ quantity, these new enormous statistics
data samples may allow us to provide comparisons of the direct $p_\mu$
and $\theta_\mu$ quantities and compare data and model in this parameter
space.  It may not be possible to explicitly provide the energy transfer $\omega$
at fixed angle as can be done for electron beams, but we should consider
that there may be a way of more directly expressing the observed lepton kinematics.
And of course, none of these distributions address uncertainties in
the rescattering of the recoil proton.

\subsection{Very low momentum transfer}

Finally, I briefly mention the never understood very low $Q^2 < 0.2$ (GeV/c)$^2$ region.
This has historically been an area of great model uncertainty, and a difficult
one to address through electron scattering experiments.
Difficulties in reproducing the data distributions
may be related to an intrinsic problem with the $Q^2$ spectrum,
or partially due to a distortion of the reconstructed $Q^2$ spectrum.
This part of the spectrum corresponds to forward
going muons in these experiments, where traditionally the detection 
efficiency is outstandingly high and uniform, but where the angular resolution of
the detectors may play a role.
Because it figures prominently in the MiniBooNE oscillation analysis,
and because they have presented a parameterization to account for it,
it has received renewed attention.  The discussion in the coming years
will likely lead to more suggestions for the cause of this, 
and ideas about how to extract information from the current and upcoming
neutrino data.


\begin{theacknowledgments}
  The author appreciates numerous discussions
and feedback from
members of the K2K, MiniBooNE, MINOS, and MINERvA experiments
and especially some who participated in the bubble chamber measurements.
I'm grateful for others for several figures that appeared in my talk
and in particular thank Sam Zeller for the journal-style plot of 
the QE measurements used here.  I also thank Hugh Gallagher and Gaku Mitsuka
for providing event samples for the {\sc neugen} and {\sc neut} comparisons 
presented here.
The event generators were run by the generator authors using a version
that was current in the Winter of 2007 to provide a comparison
for the NuInt07 conference.  Some changes may have occurred
between the description given in the citations below and what
was provided for the comparisons here.

\end{theacknowledgments}


\end{document}